# Jokeasy: Exploring Human-AI Collaboration in Thematic Joke Generation

Ge, Yate[a]; Tian, Lin[a]; Xu, Chiqian[a];Xu, Luyao[a]; Li, Meiying[a]; Hu, Yuanda[a]; Guo, Weiwei*[a]

[a] College of Design and Innovation, Tongji University, Shanghai, China
* weiweiguo@tongji.edu.cn

Thematic jokes are central to stand-up comedy, sitcoms, and public speaking, where contexts and punchlines rely on fresh material—news, anecdotes, and cultural references that resonate with the audience. Recent advances in Large Language Models (LLMs) have enabled interactive joke generation through conversational interfaces. Although LLMs enable interactive joke generation, ordinary conversational interfaces seldom give creators enough agency, control, or timely access to such source material for constructing context and punchlines. We designed Jokeasy, a search-enabled prototype system that integrates a dual-role LLM agent acting as both a material scout and a prototype writer to support human–AI collaboration in thematic joke writing. Jokeasy provides a visual canvas in which retrieved web content is organized into editable inspiration blocks and developed through a multistage workflow. A qualitative study with 13 hobbyists and 5 expert participants (including professional comedians and HCI/AI specialists) showed that weaving real-time web material into this structured workflow enriches ideation and preserves author agency, while also revealing needs for finer search control, tighter chat–canvas integration, and more flexible visual editing. These insights refine our understanding of AI-assisted humour writing and guide future creative-writing tools.

## 1   Introduction

Thematic jokes are central to stand-up comedy, sitcoms, and humorous public speaking, where punchlines rely on fresh, culturally relevant material—news, anecdotes, and trends that resonate with a specific audience(Dean & Allen, 2000; Mintz, 1985). This makes thematic joke writing an inherently context-driven and dynamic creative process, one that unfolds in three intertwined steps: (1) material gathering—hunting for timely stories, memes, and references; (2) inspiration framing—distilling those findings into angles and narrative evidence that can support a setup; and (3) joke crafting—weaving a concise setup-punchline structure that capitalizes on the collected context. While recent advances in Large Language Models (LLMs) have shown significant potential for generating humor(Gao et al., 2023; Gero et al., 2022; Tikhonov & Shtykovskiy, 2024), they often struggle to meet these dynamic knowledge requirements. Relying on static training data, LLMs can produce stereotyped or bland content that lacks the specific, timely context needed for effective humor.

Human–AI co-writing systems can lower the entry barrier for joke writing and inspire a wider range of ideas. Coupling such systems with search tools is particularly promising because retrieval-augmented LLMs can mitigate hallucinations (Shuster et al., 2021), improve factual accuracy (Feldman et al., 2024), and provide fresh, culturally aligned material (Xiong et al., 2024). These qualities help thematic jokes resonate more strongly with their intended audiences. However, we still lack a clear understanding of how human writers and search-enabled LLM agents can collaborate effectively throughout the process of thematic joke creation. Open questions include how real-time, web-sourced information can be presented and integrated in human–AI co-creation to enrich rather than overwhelm the creative process, how responsibilities should be divided between the human writer and the AI during ideation, refinement, and prototype generation, and what types of interfaces can best reveal the agent's reasoning while maintaining user control and creative flow.

To explore these open questions, we adopted a research-through-design approach. Jokeasy is a functional prototype system developed to examine how humans and AI collaborate in thematic joke writing. It serves as a working research platform that enables observation, reflection, and analysis of human–AI interaction in the creative process. After implementing the prototype system, which integrates search-enabled human–LLM co-writing and supports both real-time material scouting and joke prototype drafting, we conducted a user study with professional and hobbyist comedians. We recorded their on-screen actions, think-aloud comments, and post-session interviews, followed by thematic analysis. Treating Jokeasy as a functional prototype allowed us to focus on how writers negotiated agency, reused retrieved information, and navigated the visual workflow during collaboration with the AI partner.

This paper makes the following contributions:

- **Jokeasy, a prototype system.** We present a search-enabled visual co-writing system that allows human authors and an LLM agent to gather web material, organize inspiration blocks, and create thematic joke prototypes collaboratively.
- **Empirical insights into human–AI collaboration in thematic joke writing.** A qualitative study with 13 hobbyist writers and 5 experts (professional comedians and HCI/AI specialists) illuminates how creators negotiate agency, reuse retrieved web material, and navigate a multistage visual workflow when co-authoring humour with the agent.
- **Design implications for future humor-writing tools.** The study distils three design insights and offers preliminary recommendations for future AI-assisted humour writing and broader human–AI collaborative creative-writing tools.

## 2   Related Work and Design Rationale

### 2.1 Human-AI Co-Creation in Creative Writing

Recent works in AI-assisted creative writing have explored how LLMs can support ideation during the creative process. For example, AngleKindling (Petridis et al., 2023) helps journalists generate diverse prototype by using LLMs to ideate angles from press releases, significantly reducing the mental load for idea generation. Similarly, AI-Augmented Brainwriting (Shaer et al., 2024) incorporates LLMs into the ideation and evaluation stages, improving the quality of group-based ideation in creative contexts. Other works like Jamplate (Xu et al., 2024) and Sparks (Gero et al., 2022) leverage LLMs to support reflection and refinement in science writing by integrating structured templates and dynamic prompts to stimulate idea generation.

Information search plays an indispensable role in creative support systems, where relevant and timely results significantly enhance the quality of generated content. Some recent LLM-based creative support systems are equipped with integrated search tools, improving various aspects of the creative generation process, such as sourcing multiple inspirations (Chavula et al., 2023), interpreting intermediate results (Radensky et al., 2025) and improving the quality of creative generated content (Shi et al., 2023). However, these systems are seldom applied to writing tasks that depend heavily on up-to-date cultural and social contexts—such as humor and joke creation. Few existing tools consider the AI not only as a co-writer but also as an active material scout.

Our work addresses this gap by examining how real-time, search-enabled LLMs can better support human–AI collaboration, particularly in joke writing, by empowering users through contextual relevance, evidence grounding, and iterative creative control.

### 2.2 Design Interfaces for LLM-Based Creativity Support System

Compared to traditional chat-based collaborative creativity support systems, structured visual representations and visual interfaces better facilitate human-AI collaboration in creative tasks[20]. Methods in prior studies, such as node-based (Jiang et al., 2023; J. Kim et al., 2023), modular (T. S. Kim et al., 2023; Ma et al., 2024), and "fragment-based collage"(Buschek, 2024; Reza et al., 2024), are emerging as a trend in LLM-based creativity support contexts. These methods allow users to construct inputs using basic creative units to "prompt" system responses.

Furthermore, continuous multi-turn communication between the user and the system is crucial for refining the creativity support content. Structured design interfaces enable users to perform structured editing. VISAR (Zhang et al., 2023) allows users to create and edit node-based text fragments via a drag-and-drop interface, adjusting the argument structure or content by modifying nodes or connections. To accommodate diverse human-AI communication needs, some systems offer multiple interaction modes. For instance, Sensecape (Suh et al., 2023) provides canvas and hierarchy views, enabling users to explore and construct knowledge structures at different abstraction levels using structured representations of outputs. LMCanvas (Kim et al., 2023) breaks text into actionable objects, allowing users to organize tasks and manage writing variants according to their preferences.

While prior work on structured visual interfaces has primarily focused on reflective or argumentative writing (e.g., VISAR, LMCanvas, ABScribe), these systems do not address creative processes that depend on real-time cultural context. Jokeasy extends this line of work by integrating a search-enabled workflow into a visual co-writing environment. It allows users to retrieve topical material, organize it into editable inspiration blocks, and iteratively transform it into joke prototypes. This combination of retrieval, visualization, and humor-specific workflow distinguishes Jokeasy from previous creativity-support systems and positions it as a new form of search-grounded, visual co-authoring interface.

**2.3 Thematic Joke Writing**

Thematic joke writing—humor centered on a specific topic, current event, or cultural reference—has gained wide popularity among everyday humor enthusiasts who occasionally engage in joke creation (Dynel, 2016). On platforms like Reddit, Xiaohongshu, and open mic circuits, creators frequently use this genre to build jokes that feel timely and relatable. The typical creative workflow starts with sourcing inspiration from recent news headlines, social media memes, or personal anecdotes, followed by divergent ideation that explores multiple humorous angles. This practice is inherently improvisational and iterative: creators often remix and refine their material through trial and error until the comedic rhythm and relevance align with audience expectations.

This process can be decomposed into two interdependent challenges.

First, thematic joke writing involves distinct creative demands that set it apart from general text generation. Rather than producing language alone, joke creation requires careful cultural awareness, and timely content sourcing. The success of a thematic joke often hinges on how well it integrates real-world references—news, memes, lived experiences—into a narrative form that sets up and subverts expectations (Attardo, 2001; Chen et al., 2024). Punchlines must resonate with audience knowledge, humor norms, and shared context.

Second, thematic joke writing involves both structural and cognitive complexity that challenges conventional, linear writing workflows. Structurally, jokes require a setup-punchline architecture that must unfold logically, often under strict timing constraints. Cognitively, creators frequently explore multiple angles, modify previous attempts, and remix fragments to reach novelty and surprise (Frich et al., 2021).

These constraints expose the limitations of conventional chat-based LLM interfaces. Designing for thematic humor creation thus demands visual, structured, editable, and context-aware interfaces.

# 3  Jokeasy: System Design and Implementation

This section details the design and implementation of Jokeasy, a prototype system developed to be studied through qualitative methods, aiming to uncover how users collaborate with an AI agent to craft fresh, audience-resonant thematic jokes.

## 3.1  Design Considerations

To use Jokeasy as a prototype system for studying human–AI collaboration in thematic joke writing, we defined three interrelated design considerations that guided the development of both the LLM agent logic and the user interface.

- **DC1. Dual role for the LLM agent: prototype writer ＋ material scout**

    Because thematic joke writing depends on fresh and timely references, our system asks the LLM to do more than simply draft jokes. It first operates as a material scout, searching the web according to the user's stated theme and intent for memes, headlines, and anecdotes that can support later collaboration. Then, using the retrieved web content, it acts as a prototype writer that assembles a setup-and-punchline skeleton for the writer to refine. This combination allows the AI to provide breadth through up-to-date material, while the human creator contributes depth by selecting the references that best fit their voice and audience..

- **DC2. Multi-stage collaboration workflow built on inspiration blocks derived from search results**

    Extending DC 1, Jokeasy turns each retrieved search result into a concise inspiration block—a manipulable unit that the human writer and the LLM can build on instead of injecting raw text directly into a joke. The collaboration unfolds in four sequential steps: (1) **topic ideation**—specifying the theme, audience, and stylistic cues; (2) **inspiration generation and initial prototype creation**—the agent proposes candidate angles, contextual facts, and drafts initial jokes grounded in those cues; (3) **inspiration validation and collaborative refinement**—the writer reviews and evaluates the AI-generated content, and further refines the inspiration blocks independently or in collaboration with the AI; and (4) **final joke synthesis**—after ongoing collaborative editing between the user and the AI, the system produces the final draft of the joke prototype. Using inspiration blocks as a shared creative currency keeps intent aligned, supports ongoing revision, and lets the AI contribute breadth while the human maintains creative depth.

- **DC3. Visual, object-based canvas to externalise the conversation**
    Building on DC1's dual-role agent and DC2's four-stage workflow, Jokeasy uses a visual canvas to present everything co-produced by the agent and the writer, including search-derived material and evolving joke ideas, within a single, easily viewable workspace. By transforming

these otherwise hidden contributions into tangible elements, the canvas allows writers to inspect, edit, and reorganize their collaboration at any time, maintaining alignment across stages and preserving creative agency.

## 3.2 System overview

We designed and developed the Jokeasy system (as shown in Figure 1), which allows users to interact with a search-enabled LLM agent through a visual interface to generate multiple joke prototypes based on a given theme. The Jokeasy system divides the joke creation process into four phases, as defined by DC2.

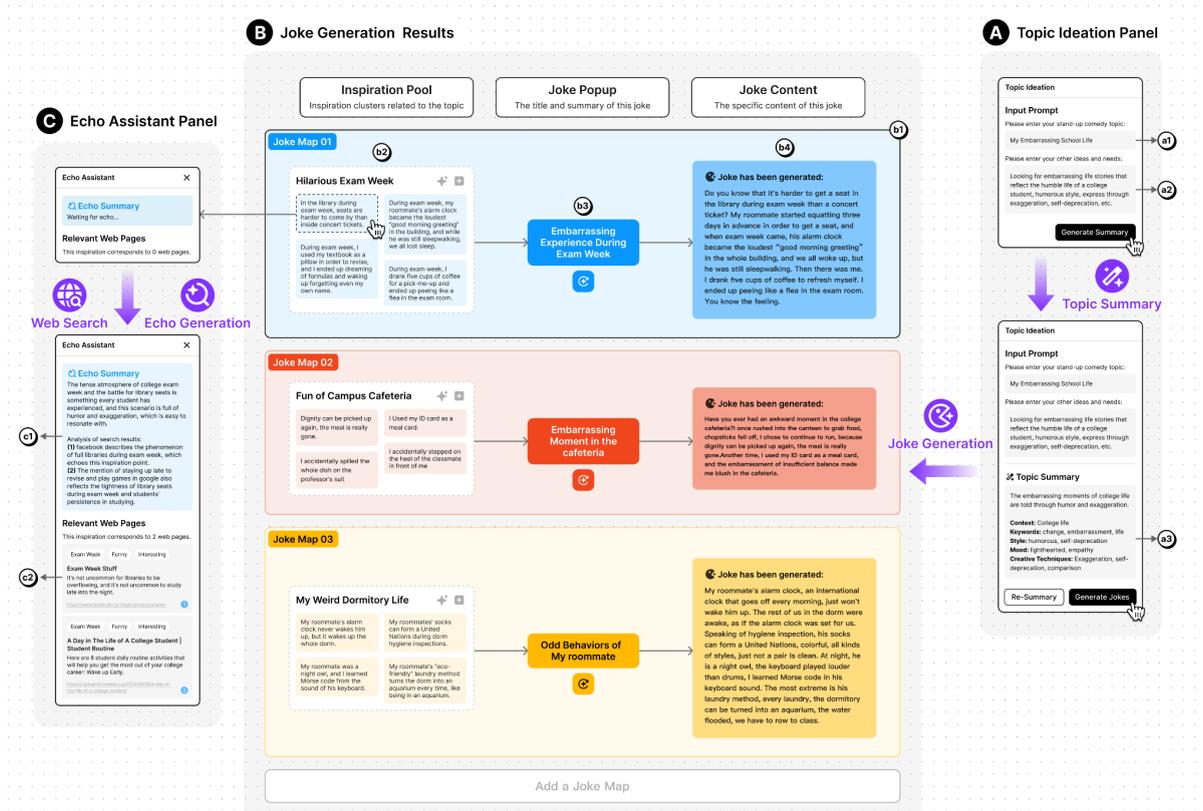

*Figure 1. The main interface of Jokeasy.*

**Topic Ideation Phase**

The user begins by dragging a Topic Ideation Panel (Fig. 1 A) onto the canvas, which opens an input form for the new routine. Here, they enter the creation topic along with any supplementary ideas and requirements—such as expected scenarios, preferred styles, and specific comedic techniques. Jokeasy then analyses this input and generates a structured topic summary for review. If the summary does not match the writer's intent, the user can revise the input and click Re-Summary to regenerate it; once the summary aligns with their direction, pressing Generate Jokes advances the workflow to the next stage.

**Inspiration generation and initial prototype creation**

Once the topic summary is confirmed, Jokeasy's LLM agent derives three high-level *inspiration themes* from it. For each theme the system (1) expands a set of search keywords, (2) queries the web for timely, relevant material, and (3) distils that material into concise *inspiration blocks* that populate an inspiration pool. The LLM then merges the topic summary with the blocks in each pool to produce a provisional joke title and prototype joke content, together forming a joke map (Fig. 1 b1).

On the canvas, Jokeasy presents three such maps side by side. Each map shows its inspiration pool, title, and draft joke. Selecting an inspiration block opens the **Echo Assistant** (Fig. 1 C), which displays the retrieved source material and an echo summary explaining how the block could resonate with the target audience—information that guides refinement in the next phase.

**Inspiration validation and collaborative refinement**

During validation, the writer inspects and refines the contents of each joke map. Selecting an inspiration block opens the Echo Assistant, which displays the retrieved source material together with an echo summary that explains how that material could resonate with the intended audience. If a block seems weak, off-topic, or if the writer wishes to develop a new angle, they can edit or delete it—every change prompts Jokeasy to rerun the web-search step and regenerate the associated echo summary so that evidence stays current (Fig. 2 A). The writer can also expand the inspiration pool within the same map: choosing *AI Add* asks the agent to search again under the current theme and insert a fresh block, whereas *Manual Add* lets the writer draft a new block first, after which the system retrieves supporting information and appends a corresponding echo summary.

When broader reorganisation is needed, the writer can introduce additional joke maps on the canvas—or remove any map that is no longer relevant(Fig. 2 B). In AI-generated mode, Jokeasy reruns the full pipeline—theme extraction, web retrieval, inspiration-block creation, and initial prototype drafting—so the new map appears fully populated and ready for review. In manual mode, an empty map frame is added; the writer may sketch their own inspiration blocks and invoke the agent later to fetch material and complete the title and prototype joke.

**Final joke synthesis**

After the user has continuously reviewed and refined the generated joke content with the LLM agent, and confirmed that it aligns with their creative intent, the collaborative creation process for the thematic joke is completed.

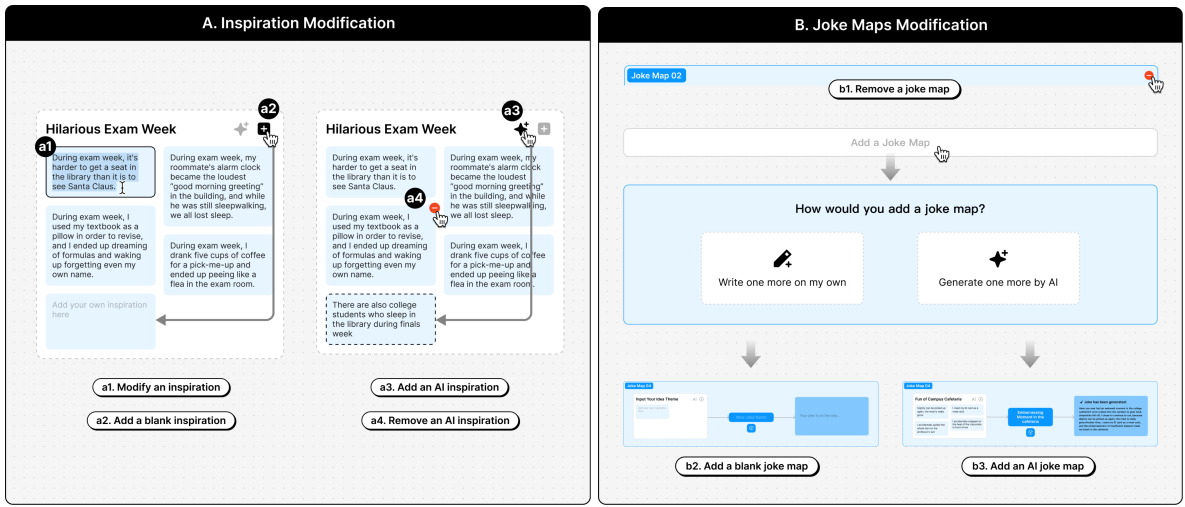

*Figure 2. Jokeasy's interface about inspiration modification and joke map modification.*

### 3.3 System Implementation

The front-end of Jokeasy is developed as a Figma widget plugin. The back-end is built using Node.js. The large language model we use is moonshot-v1-auto 3 , which supports up to 128k tokens. The model's temperature parameter is set to 0.3. As shown in Figure.3, we implemented the core LLM-powered functions of the Jokeasy system using the LLM-chain method and the structured output feature. The prompt preamble in our system is organized into six key fields: [Role], [Input Context], [Overall Rules], [Output Formatting], [Workflow], and [Example] (examples shown in Figure.4 ). This structured approach helps to build LLM modules in Jokeasy's LLM pipeline to be able to accomplish various tasks. The search functionality is implemented using the Tavily API .

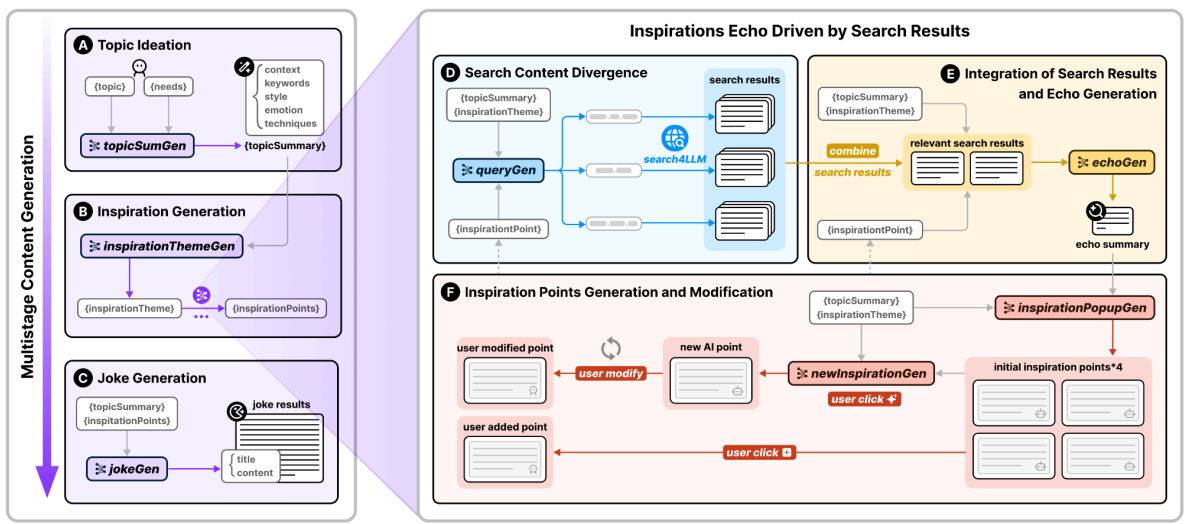

*Figure 3. The LLM-based pipeline of the Jokeasy system.*

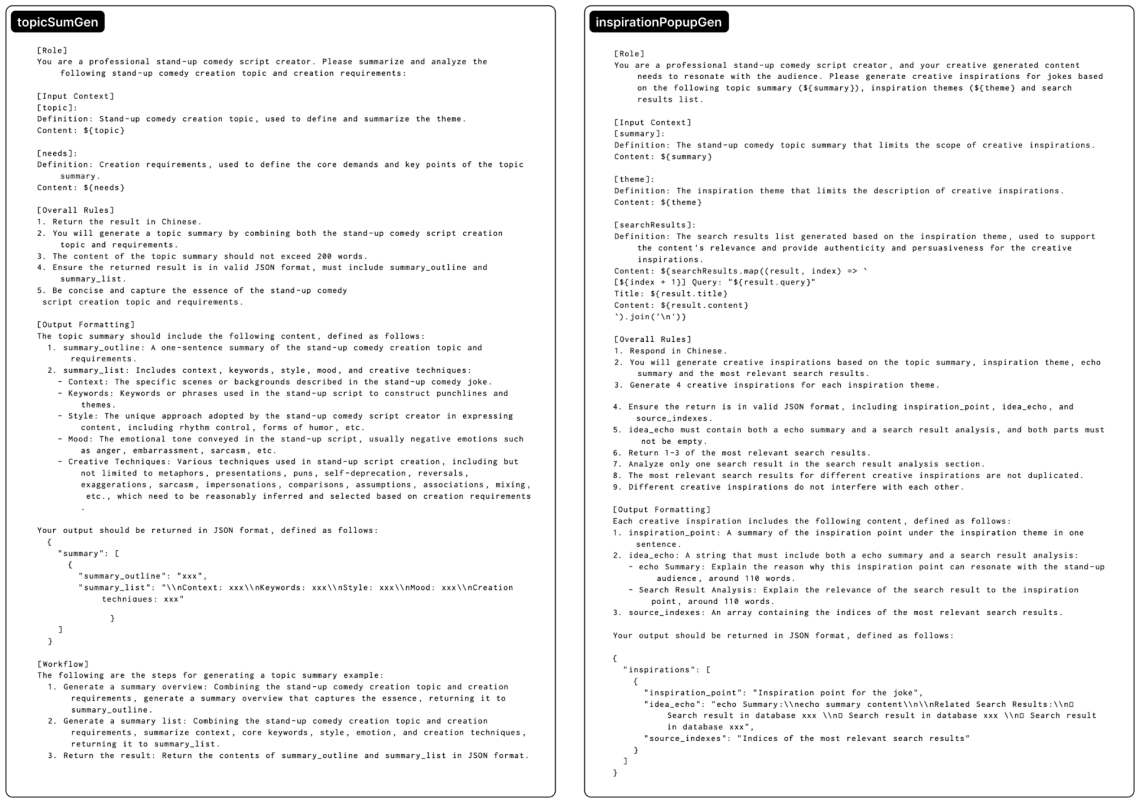

*Figure 4. LLM prompt templates: topicSumGen and inspirationPopupGen.*

## 4 User Study

### 4.1 Participants

We recruited 18 participants (P1–P18) for the study. General users (P1–P13), comprising hobbyists and enthusiasts, were recruited via an online questionnaire distributed through social media platforms. Expert users (P14–P18, referred to as E1–E5) were directly invited. Among them, E1–E3 were comedians with over five stand-up performance experiences. E4 and E5, passionate stand-up comedy enthusiasts, brought additional expertise: E4 is a human-computer interaction specialist, and E5 is an AI researcher. The expert interviews aimed to gain in-depth insights into the technical feasibility, design implications, and broader applications of the Jokeasy system.

The general users received 40 RMB as compensation for their participation, while the expert users were compensated with 200 RMB. Each session lasted approximately one hour.

### 4.2 Procedure

The study consisted of three phases: pre-test, task execution, and a post-test interview. Upon arrival, participants provided informed consent and completed a pre-test questionnaire to collect demographic information, stand-up comedy writing experience, familiarity with AI-assisted writing tools, and challenges they typically encounter when using LLM-based tools. Participants were then briefed on the overall structure of the study and provided with a list of creative tasks.

During the testing phase, participants individually interacted with the two systems (the baseline system (Figure 5), and the Jokeasy system) in a counter-balanced order to mitigate potential order effects. The baseline system was a conversational interface that used the same LLM model as Jokeasy, with search capabilities and the ability to integrate retrieved information into its outputs, but without Jokeasy's multi-step search integration, staged workflow, or visual co-creation features. Before testing each system, participants watched a short video tutorial to familiarize themselves with its functionality.

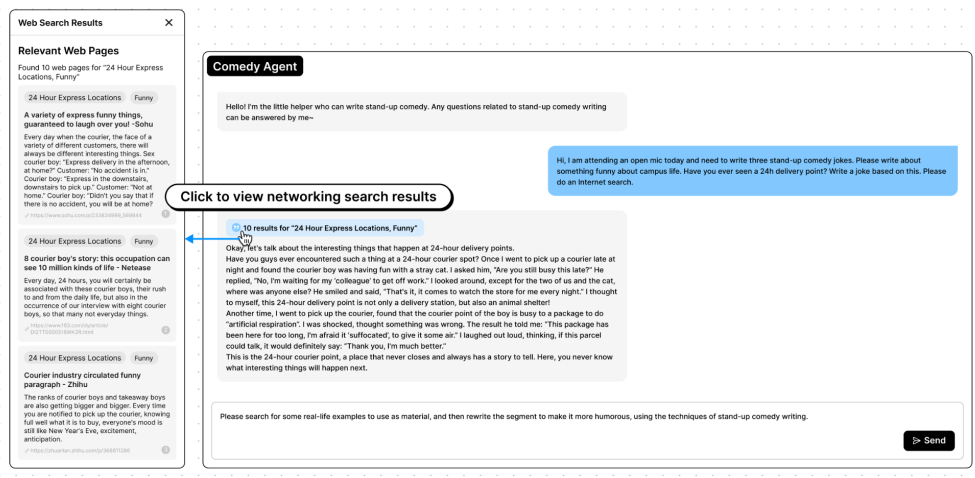

*Figure 5. Interface of the baseline system.*

For each system, participants were asked to complete a stand-up comedy joke creation task, with a suggested time limit of 15 minutes per task, though no strict time constraints were imposed. During this phase, participants were encouraged to verbalize their thoughts and decisions using a think-aloud protocol, enabling the collection of qualitative insights. Their interactions, including on-screen actions and spoken thoughts, were recorded via video and system logs for later analysis.

Once both tasks were completed, participants took part in a 15-minute semi-structured interview. This session allowed them to reflect on their overall experience with the two systems, discuss perceived strengths and limitations, and provide suggestions for improvement. This procedure ensured comprehensive qualitative data collection to evaluate the effectiveness and usability of the systems.

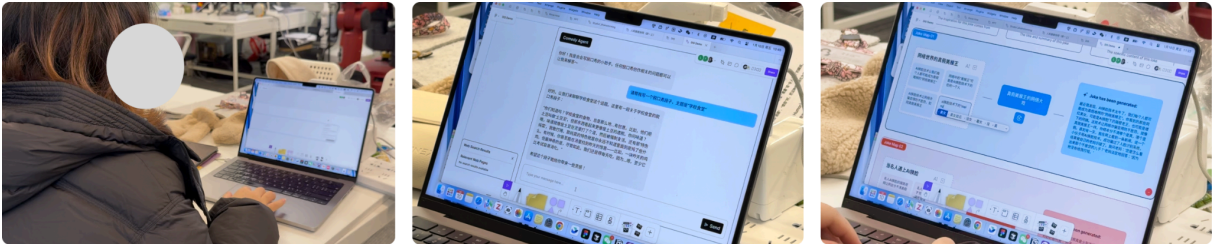

*Figure 6. One of the user study scenarios. All participants were native Chinese speakers, and LLM outputs were configured to be in Chinese.*

## 5  Findings

### 5.1  Overall experience

Most participants (13 / 18) favoured Jokeasy over the baseline chat tool for thematic-joke creation. They described Jokeasy's four-stage workflow as "organised and sequential from inspiration to the final product" (P10) and felt it "integrated several small functions involved in joke writing" (P5). By contrast, the baseline left them "uncertain about the system's understanding of my needs, making the generated results feel like opening a blind box" (P14). Participants also liked the additional flexibility Jokeasy offered: it "provides options for adding, deleting, and modifying" (P7) and thereby "lets me focus on creativity and refining details" (P3) instead of constant re-prompting. As P13 added, Jokeasy gives "great freedom in editing." Even so, three users still saw Jokeasy more as a content generator than a true collaborator; one remarked that the conversational baseline "gave me a stronger sense of collaboration because jokes are developed through conversation, which is closer to actual stand-up performance" (P2).

### 5.2  User Perception on human-AI Collaboration in Jokeasy

**Search-driven inspiration and the AI's dual role**

Eight participants said the agent reliably matched their intentions, "continuously sparking new ideas" (P6), "coming up with angles I hadn't thought of" (P18), and even "inspiring me when I ran out of ideas" (P9). Integrated search helped them uncover "the real emotions behind this inspiration point" (P7) and "unexpectedly get new ideas while browsing results" (P1). The one-to-one link between each inspiration block and its supporting web evidence let users "browse more efficiently and easily find key information" (P8) and "trace back why the AI generated this inspiration" (P18). Still, some hits were judged "too general" (P16) or "not strongly related to the inspiration point" (P17). Others felt the system occasionally "integrated too many ideas into the jokes, neglecting specific and authentic content descriptions" (P5) or that "initial broad demands led to unnecessary redundant content" (P12).

**Visual canvas as a collaboration scaffold**

Fourteen participants praised the canvas for its structured clarity; it "has a clear framework, and the logical structure of the visual interface helps me break down and organise the narrative, from theme to style to joke connection" (P1). Several compared it to a mind map: "The process, which leans towards mind-map recording, is clear" (P12); the card layout "is very flexible, corresponding to my fragmented ideas" (P15). They enjoyed being able to "create many maps and let many ideas develop concurrently" (P9); the parallel maps "provide a sense of contrast, like three different results for the same material" (P1). Editing—whether adding or revising cards—was "relatively convenient" (P10) and "easy to modify" (P11, P13). A few participants noted that the interface felt "slightly complex at first" (P2, P4, P13), but believed it would "effectively assist the creative process in the long run" (P4).

## 5.3 Use case

To better understand how users collaborate with AI systems in the creative process of thematic joke writing, we present two real-world interaction scenarios using P7 as a representative participant.

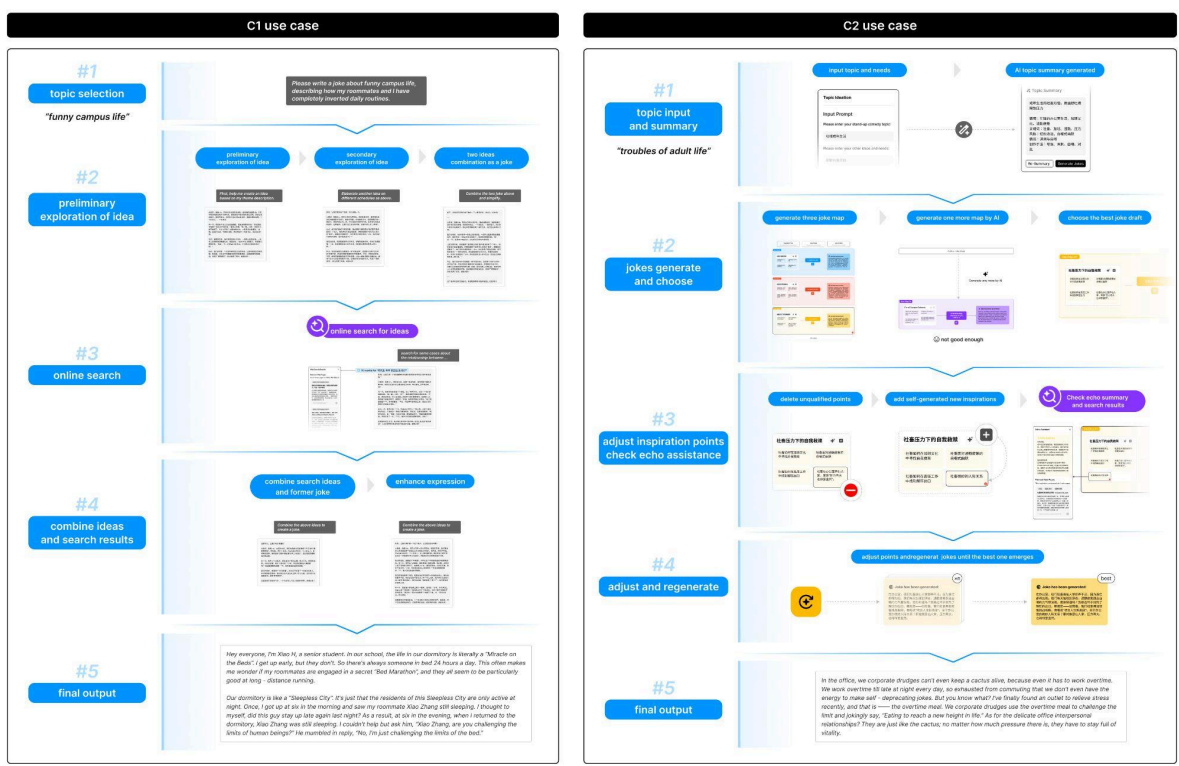

Figure 7. Step-by-step real-world interactions of P7 with the baseline and Jokeasy systems.

**C1. Baseline Use Case**

P7 began by selecting the topic "**Funny Campus Life**" and entered the prompt: *"Please give me an idea for joke writing about funny campus life, describing how my roommates and I have completely inverted daily routines."* The baseline system initially returned a basic idea based on the prompt. To explore alternative perspectives, P7 requested a second version with a different comedic angle. After reviewing both outputs, he asked the system to merge the two ideas. In response, the baseline system generated a third version that integrated elements from both drafts into a coherent joke.

P7 expressed dissatisfaction with the joke and proactively modified the prompt by adding a request for online search. He said that he was hoping to find real-life cases related to roommate relationships as new sources of inspiration. After a short while, the baseline returned ten relevant cases and provided corresponding idea summaries within the dialogue flow.

P7 asked to combine parts of the initial joke with insights from the search results. He subsequently asked to "*enhance expression*" and finally got the preferred joke content.

**C2.Jokeasy Use Case**

First, P7 entered the topic "**Troubles of Adult Life**" in the Topic Ideation panel and specified creative needs, including an emphasis on exaggerated expressions and a focus on subtopics such as workplace burnout. Upon clicking "Generate Summary," Jokeasy produced a structured summary that met P7's expectations on the first attempt.

Subsequently, P7 clicked "Generate Jokes." Jokeasy then generated three distinct joke maps derived from an Inspiration Pool of AI-generated ideas, including examples such as "*The Absurd Theater of Overtime Nights*." P7 proceeded to click "Add a Joke Map" and opted to have the AI generate an additional one. With four joke maps now available, P7 reviewed them in sequence and selected Joke Map 03, titled "*Self-Rescue Under Work Pressure*," for further refinement.

To tailor the content, P7 removed one inspiration point that was not considered engaging. He also added a blank inspiration point and manually input an idea he found personally relevant: the subtle dynamics between colleagues. By selecting this point, P7 accessed the Echo Assistant panel, where he reviewed the corresponding echo summary and a list of relevant web pages. One of the pages was opened for closer inspection.

Following this, P7 used the "Regenerate Joke" function three times, reviewing each version in turn. He ultimately selected the output he found most aligned with his intent as the preferred joke content.

### 5.4 User expectations of Jokeasy's real-world application, system functionality and interaction quality.

Most participants believed that Jokeasy could extend beyond **thematic-joke creation** and support other writing scenarios. For example, several noted its potential for speech drafting or debate scripting (5 / 18), while others mentioned literature reviews or workplace summaries (4 / 18). In addition, participants envisioned using Jokeasy for information-organisation tasks such as mind-map knowledge bases (P7), class notebooks (P1), and academic-research (P8).

Participants also offered concrete improvement ideas. P8 and P9 suggested integrating the visual canvas with a conversational side-panel; five participants asked for a history log to revisit previous iterations; and P14 wanted a "favourite lines" repository for standout jokes. P11 and P12 recommended better prioritisation and sorting of inspiration points, whereas P8 and P17 called for an in-interface guidance mechanism. Finally, P2 and P4 wanted the transition from the Topic Ideation Phase to the first Inspiration Generation Phase to be more configurable, allowing users to set how many inspiration points and prototype jokes are produced at that hand-off, so they retain greater control over the creative space they wish to explore.

## 5.5 Expert Interview

Unlike the general user sessions that centred on day-to-day co-authoring, the expert interviews focused on Jokeasy's practical value and future role in professional humour practice.

**Overall attitude**

Experts agreed that the system offers strong support for novices—"Jokeasy suits new comedians lacking systematic training" (E2)—and helps experienced writers diversify topics (E3). They hope later versions will surface specific comedic techniques (E1, E4). While they value AI as a drafting aid, they doubt it can fully "understand" humour; as E3 put it, "AI can generate useful references, but final judgement must come from human creators." Some also worried that easy AI generation might dilute originality and threaten professional livelihoods (E3).

**Technical potential and current limits.**

All five experts saw clear value in pairing an LLM with live search, especially for humour that must reference current events. E5 noted, "Leveraging RAG to pull external knowledge gives richer inspiration and helps integrate news into jokes." Yet they pointed to weaknesses: narrow sources (E5), overly broad or low-value hits (E1, E3), and controversial items (E3) that users must filter. E3 stressed that web search "works best for attitudes, opinions, real events," while E5 suggested switching to a deeper search API. Future improvements could include fine-tuning on comedy corpora (E5) and building a personalised reference library (E4).

**Search-enabled collaboration**

Four experts emphasised that search makes the tool more practical. E3 observed that "checking existing evaluations and jokes on a topic provides valuable inspiration," and search lets writers "capitalise on trends." Results for a single inspiration block can repeat; E5 proposed querying from several viewpoints (public opinion, news, expert knowledge). Echo summaries were seen as a quick resonance check (E1, E4) but only a "rough estimate" (E2); E1 suggested highlighting negative emotions because "that's core to stand-up."

**Visual canvas**

Most experts (4 / 5) felt the non-linear canvas fits joke-writing practice. E4 said, "The process starts with isolated inspiration cards that grow into a larger map," and users can trace blocks easily (E3). Still, they want more room for mid-stage edits: "Writing a joke is digging deeper into a point; I prefer mind-map editing" (E1, E3). Three experts argued that conversational interaction feels more natural for comedians, and E5 noted that the canvas "limits flexibility; users can't freely decide what to search or how to reuse results."

# 6 Discussion

Our findings from the user study, particularly the comparison between the visual-centric JOKEASY and a baseline conversational system, offer key insights for designing future AI-assisted creative tools. We frame our discussion around our three core design principles.

## 6.1 Design Implications for AI-assisted Thematic Joke Writing

Our findings from deploying Jokeasy as a prototype system highlight several important considerations regarding its current design. Users particularly appreciated Jokeasy's clear role distinction between the AI as a "material scout" and a "prototype writer." This dual role allowed authors to benefit from timely, culturally relevant inspirations while maintaining sufficient control to tailor content to their audience and creative intent. However, users indicated that the search results occasionally lacked precision or were overly general, prompting requests for more explicit control over source selection to improve relevance and specificity.

The structured visual workflow provided notable strengths by clearly organizing the creative process into distinct stages, from topic ideation to joke synthesis. This organization made creative development more systematic and transparent. Participants reported that this structure significantly reduced cognitive load and allowed easy tracking and editing of ideas. However, this approach also introduced a rigidity that some users found restrictive, limiting spontaneity and improvisation, which are qualities often essential to creative humor writing. To address this limitation, users suggested integrating conversational elements directly into the visual canvas to enable real-time, dynamic interactions that combine the advantages of structured visualization and conversational fluidity.

Additionally, the visual canvas's editing capabilities were appreciated for their ease and clarity, facilitating straightforward idea manipulation. However, participants expressed a desire for richer interaction features, such as the ability to link, merge, or rearrange elements more freely, in order to better support iterative creativity and complex joke structures. Some participants also requested lightweight scripting or logic tools within the canvas, indicating the need for a more dynamic visual workspace that can adapt to varied creative workflows.

Drawing upon these insights, we propose that future AI-supported creative writing tools should:

1. **Configurable search support.** Provide lightweight controls, such as source toggles, depth sliders, or a quick chat option, so that writers can decide where and how widely the agent searches. Customizable retrieval keeps jokes topical while preserving the author's voice and creative agency.
2. **Canvas-Chat Integration.** Couple the structured canvas with a conversational side-panel that stays in lock-step. Chat supports spontaneous riffing; the canvas records decisions and evidence, acting as an objective shared memory.
3. **Expanded visual editing.** Extend the canvas with drag-to-link, merge, and free-form rearrangement features to let writers remix blocks mid-process. This enhancement would transform the static board into a genuine visual programming space for joke construction, addressing both novices' need for guidance and experts' demand for flexibility.

## 6.2 Insight into AI for Humor Generation

Our findings reveal several humor-specific dynamics that differentiate joke writing from other creative co-writing domains. Participants highlighted the importance of temporal relevance and cultural resonance in producing effective humor. Real-time retrieval was especially valuable for generating contextually grounded jokes that connect with current social trends. Expert comedians emphasized the need to balance surprise and social appropriateness, noting that humor often relies on subtle timing, audience awareness, and sensitivity to shared cultural norms.

These observations suggest that AI systems for humor generation should not only provide linguistic creativity but also support contextual alignment and tone calibration. Future systems could include adaptive retrieval filters, audience-aware humor framing, or emotional tone adjustment mechanisms to ensure that AI-assisted jokes remain both relevant and appropriate.

## 6.3 Limitation of the study

This study has several limitations. First, the Jokeasy prototype was developed as a research-oriented system to explore early-stage thematic joke creation rather than the full cycle of comedy writing. Thus, it does not yet support comprehensive editing or refinement of complete stand-up scripts. Second, our current implementation relies on a simplified web search tool, which occasionally yielded overly broad or insufficiently specific results, potentially constraining the quality and depth of retrieved inspirations. Finally, while our qualitative user study involved a diverse group of hobbyist and expert participants, the relatively modest sample size limits broader generalizability of our findings, and the general-purpose orientation of Jokeasy may have constrained deeper insights tailored to specialized user groups or professional contexts.

For future work, we plan to extend Jokeasy by integrating conversational interactions with the visual interface to better balance structured guidance with improvisational flexibility. Additionally, we aim to enhance the search mechanism by incorporating deeper retrieval techniques and customizable sourcing controls, thus aligning the system more closely with the nuanced requirements of thematic joke writing and other context-sensitive creative tasks.

# 7   Conclusion

We introduced Jokeasy, a search-enabled visual co-writing prototype system for studying human–AI collaboration in thematic joke generation. The combination of real-time web retrieval and a structured canvas enables writers to gather, organize, and develop topical material into jokes. A qualitative study with hobbyist and professional comedians demonstrated that Jokeasy provides a clear workflow, supports strong user agency, and enhances ideation. The study also identified areas for improvement, including finer search control, better integration between chat and canvas, and more flexible visual editing. These findings advance our understanding of AI-assisted humor writing and inform the design of future creative-writing tools.

**About the Authors:**

**Yate Ge:** Ph.D. candidate at the College of Design and Innovation, Tongji University. His research focuses on interaction design, human-robot interaction, human-AI interaction, and end-user programming for generative and adaptive interactive systems.

**Lin Tian:** Master's student in Interaction Design at the College of Design and Innovation, Tongji University.

**Chiqian Xu:** Master's student in Interaction Design at the College of Design and Innovation, Tongji University.

**Luyao Xu:** Master's student in Design Strategy at the College of Design and Innovation, Tongji University.

**Meiying Li:** Master's student in Interaction Design at the College of Design and Innovation, Tongji University.

**Yuanda Hu:** Ph.D. candidate at the College of Design and Innovation, Tongji University. His research interests include human-robot interaction, robot learning, computer vision, and human-computer interaction.

**Weiwei Guo:** Associate Professor at the College of Design and Innovation, Tongji University. His research integrates AI and human-centered design, focusing on multimodal perception, embodied interaction, and human–computer/robot interaction for Earth vision, home, and healthcare applications.

**Acknowledgement:** This work is supported by Tongji Innovation Design and Intelligent Manufacturing Disciplines Project, and Tongji University National AI Industry–Education Integration Innovation